\pdfoutput=1

\documentclass[twocolumn,aps,prl,showpacs,superscriptaddress]{revtex4}

\usepackage{graphicx}
\usepackage{amsmath,amssymb}
\usepackage{color}
\usepackage{epsfig}

\def\endproof{\vrule height6pt width6pt depth0pt}

\begin{document}


\title{Quantum social networks}


 \author{Ad\'an Cabello}
 \affiliation{Departamento de F\'{\i}sica Aplicada II, Universidad de
 Sevilla, E-41012 Sevilla, Spain}
 \affiliation{Department of Physics, Stockholm University, S-10691
 Stockholm, Sweden}
 \author{Lars Eirik Danielsen}
 \affiliation{Department of Informatics,
 University of Bergen, P.O. Box 7803, N-5020 Bergen, Norway}
\author{Antonio J. L\'{o}pez-Tarrida}
 \affiliation{Departamento de F\'{\i}sica Aplicada II,
 Universidad de Sevilla, E-41012 Sevilla, Spain}
\author{Jos\'{e} R. Portillo}
 \affiliation{Departamento de Matem\'{a}tica Aplicada I,
 Universidad de Sevilla, E-41012 Sevilla, Spain}
\date{\today}


\date{\today}




\begin{abstract}
We introduce a physical approach to social networks (SNs) in which each actor is characterized by a yes-no test on a physical system. This allows us to consider SNs beyond those originated by interactions based on pre-existing properties, as in a classical SN (CSN). As an example of SNs beyond CSNs, we introduce quantum SNs (QSNs) in which actor $i$ is characterized by a test of whether or not the system is in a quantum state $|\psi_i\rangle$. We show that QSNs outperform CSNs for a certain task and some graphs. We identify the simplest of these graphs and show that graphs in which QSNs outperform CSNs are increasingly frequent as the number of vertices increases. We also discuss more general SNs and identify the simplest graphs in which QSNs cannot be outperformed.
\end{abstract}



\maketitle


\section{Introduction}


Social networks (SNs) are a traditional subject of study in social sciences \cite{Scott91,WF04,Freeman06} and may be tackled from many perspectives, including complexity and dynamics \cite{AB00,AB02,Barabasi09}. Recently, they have attracted much attention after their tremendous growth through the internet. A SN is a set of people, ``actors'', with a pattern of interactions between them. In principle, there is no restriction on the nature of these interactions. In practice, in actual SNs, these interactions are based on relationships or mutual acquaintances (common interests, friendship, kinship,etc). However, to our knowledge, SNs have never been discussed on the basis of general interactions which can give rise to them. This is precisely the aim of this work.

A first observation is that, while a SN is typically described by a graph in which vertices represent actors and edges represent the result of their mutual interactions, the graph does not capture the nature of the interactions or explain why actor $i$ is linked or not to other actors. From this perspective, the graph gives an incomplete description.

In order to account for this, we will consider the following, more general, scenario. We represent each actor $i$ by a yes-no test $T_i$ on a physical system $S$ initially prepared in a state $\rho$ and with possible outcomes 1 (yes) or 0 (no). Of course, these tests must satisfy some rules so that actual SNs naturally fit within them. More importantly, these rules must allow us to consider SNs beyond classical SNs (CSNs), defined as those in which the links between actors are determined by pre-existing properties of the actors, such as, e.g., their enthusiasm for jazz.

\begin{figure}[t]
 \centerline{\includegraphics[scale=0.33]{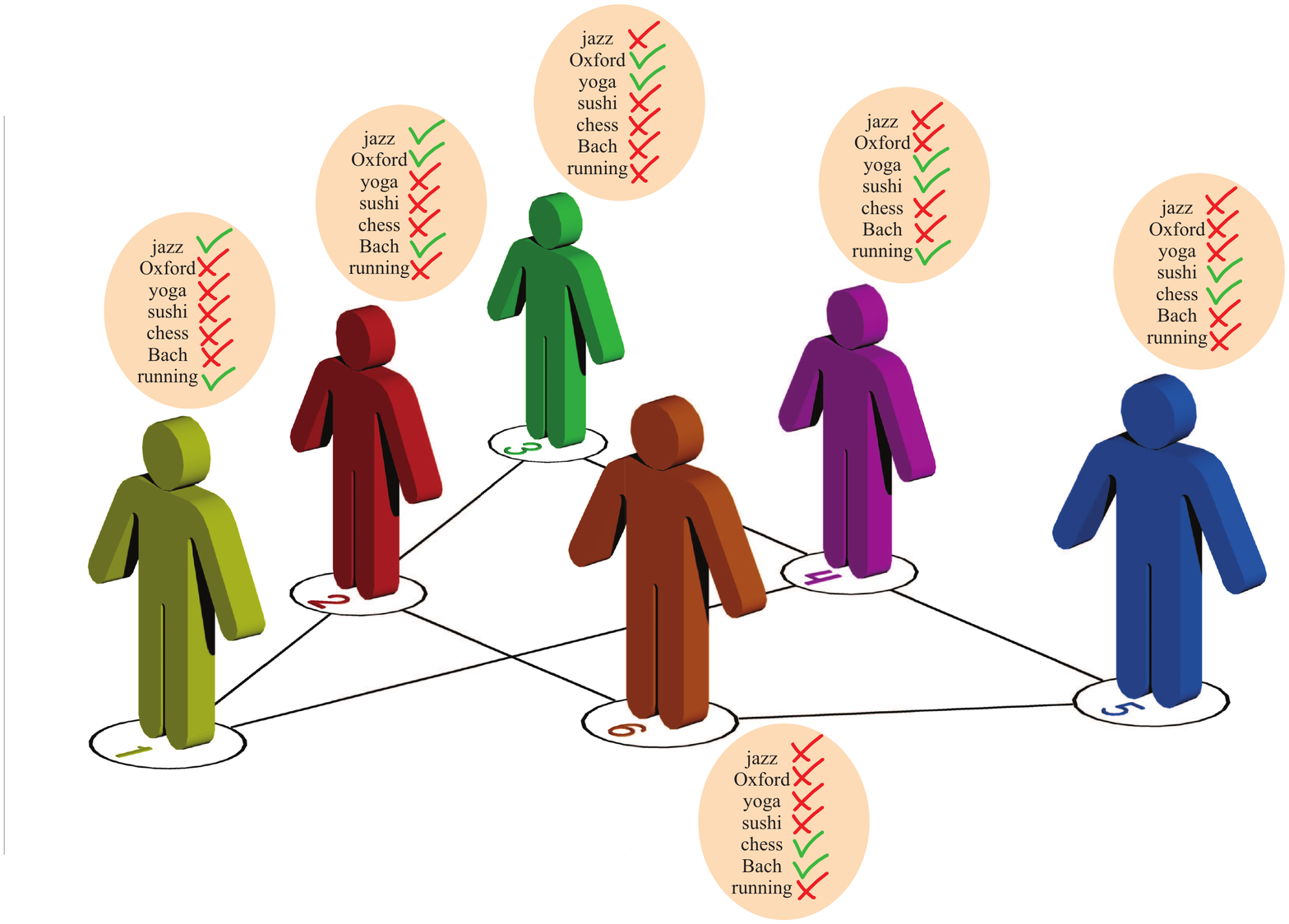}}
 \vspace{-5mm}
 \caption{A CSN. Each actor is represented by a vertex of a graph and each link by an edge.
 The characteristic of a CSN is that links can be explained on the basis of pre-existing properties
 of the actors such as
 whether or not they
 are jazz enthusiasts,
 attended Oxford University,
 practice yoga,
 like sushi,
 will participate in a chess tournament,
 love the music of J. S. Bach or
 run four times a week.}
 \label{Fig1}
\end{figure}

First, let us see how a CSN can be characterized in terms of yes-no tests $T_i$. In any CSN, it is always possible to identify a minimum set of labels such as ``jazz'' that describes the presence or absence of a link between any two actors; each actor's links are described by the value ``yes'' or ``no'' for each of these labels. An example is shown in Fig.~\ref{Fig1}. The size of that minimum set of labels coincides with a property of the underlying graph $G$, known in graph theory as its intersection number \cite{Harary94}, $i(G)$. For example, for the CSN shown in Fig.~\ref{Fig1}, the seven labels which describe the network cannot be reduced to a smaller number since for the underlying graph $i(G)=7$. Suppose each actor has the complete list of labels with their corresponding value ``yes'' or ``no'' as in Fig.~\ref{Fig1}. The input physical system $S$ can be a card, and its state $\rho$ is what is written on it, that is, the name of one of the labels. For instance, for the CSN in Fig.~\ref{Fig1}, the state $\rho$ may be ``jazz''. Then, the outcome of $T_i$ is 1 if actor $i$ has in its list of labels ``yes'' for jazz, and 0 if it has ``no'' for jazz. The initial state of the card does not change after the test.

In this characterization of a SN, tests $T_i$ naturally fulfil the following rules. (i) Two actors $i$ and $j$ are linked if and only if there exists some state $\rho$ for which the results of $T_i$ and $T_j$ are both 1, which simply means that $i$ and $j$ share the pre-existing property described by the label in $\rho$. (ii) If a test $T_i$ is repeated on the same state $\rho$, it will always give the same result, which simply means that the actor's label does not change under the execution of the test. (iii) For any $\rho$, the order of the tests is irrelevant, which reflects the symmetry of the interaction.

Note that, as a consequence of rule (i), if for a given $\rho$ the outcome of $T_i$ is 1, then a test $T_j$ in any $j$ which is not linked to $i$ will never give the outcome 1. This means that, for any $\rho$ and any set $I$ of pairwise non-linked actors, the sum of the outputs of the tests $T_i$ over all actors in $I$ is upper bounded by 1. Moreover, this results in a restriction on the possible states $\rho$. For instance, in the previous example, the state ``jazz or Oxford'' is not allowed. For such a state, the tests $T_i$ and $T_j$ corresponding to two non-linked actors $i$ and $j$ for which the values of the labels ``jazz'' and ``Oxford'' are, respectively, ``jazz: yes; Oxford: no'' and ``jazz: no; Oxford: yes'', would give both the outcome 1, in contradiction with rule (i).

Let us consider now more general SNs (GSNs). We denote by $P_{\rho}(a, b|T_{i}, T_{j})$ the joint probability of obtaining outcome $a \in \{0,1\}$ when performing $T_i$ on the system $S$ initially prepared in the state $\rho$, and outcome $b \in \{0,1\}$ when performing $T_j$ on $S$ in the state $\rho_i$ resulting from the previous test $T_i$. The tests must obey the following rules, which generalize (i)--(iii): (I) $P_{\rho}(1, 1|T_{i}, T_{j})$ determines the linkage between actors $i$ and $j$; they are linked if and only if there exists some $\rho$ such that $P_{\rho}(1, 1|T_{i}, T_{j})>0$. (II) For any $\rho$, $P_{\rho}(a, a|T_{i}, T_{i})=P_{\rho}(a|T_i)$, i.e., when $T_i$ is repeatedly performed on $S$ initially in the state $\rho$, it always yields the same result. (III) For any $\rho$, the order of the tests is irrelevant: $P_{\rho}(a, b|T_{i}, T_{j})=P_{\rho}(b, a|T_{j}, T_{i})$.

Note that in a CSN the probabilities $P_{\rho}(a|T_{i})$ can take only the values 0 or 1, whereas in a GSN these probabilities can take any values compatible with rules (I)--(III). Moreover, here we will assume that the state $\rho$ may change according to the results of the tests $T_i$.


There is a simple task which highlights the difference between a GSN and a CSN described by the same graph $G$: the average probability $\mathcal{T}$ that, for an actor $i$ chosen at random, the test $T_i$ yields the outcome $1$. The interesting point is that $\mathcal{T}$ {\em is upper bounded differently depending on the nature of the interactions defining the SN}. For the CSN, let us suppose that the card is in the state $\rho$ (e.g., jazz). Then, for an actor $i$ chosen at random, $\mathcal{T}=\frac{1}{n}\sum_{i=1}^{n} P_{\rho}(1|T_i)$, where $n$ is the number of actors. The maximum value of $\mathcal{T}$ over all possible states $\rho$ is the maximum number of actors sharing the value ``yes'' for a pre-existing property, divided by the number of actors. This corresponds to $\frac{\omega(G)}{n}$, where $\omega(G)$ is the clique number \cite{Harary94} of $G$, i.e. the number of vertices in the largest clique. Given a graph, a clique is a subset of vertices such that every pair is linked by an edge. The term ``clique'' comes from the social sciences, where social cliques are groups of people all of whom know each other \cite{LP49}. In the example of Fig.~\ref{Fig1}, the value of the clique number for the graph is $\omega(G)=2$. As a consequence, for any CSN represented by that graph, the maximum of $\mathcal{T}$ is $\frac{\omega(G)}{n}=\frac{1}{3}$.

However, for a GSN described by a graph $G$, the maximum value for $\mathcal{T}$ compatible with rules (I)--(III) is $\frac{\alpha^* (\bar{G})}{n}$, where $\bar{G}$ denotes the complement of $G$, which is the graph $\bar{G}$ on the same vertices such that two vertices of $\bar{G}$ are adjacent if and only if they are not adjacent in $G$, and $\alpha^* (\bar{G})$ is the so-called fractional packing number \cite{SU97} of $\bar{G}$, defined as $\max \sum_{i\in V} w_i$, where the maximum is taken for all $0 \leq w_i \leq 1$ and for all cliques $c_j$ of $\bar{G}$, under the restriction $\sum_{i \in c_j} w_i \leq 1$. In the example of Fig.~\ref{Fig1}, $\alpha^* (\bar{G})=\frac{5}{2}$. Hence, the maximum of $\mathcal{T}$ satisfying rules (I)--(III) is $\frac{5}{12}>\frac{1}{3}$, attainable for instance by taking $P_{\rho}(1|T_i)=\frac{1}{3}$ for $i=1, 3, 5$ and $P_{\rho}(1|T_j)=\frac{1}{2}$ for $j=2, 4, 6$.

Note that the maximum value of $\mathcal{T}$ does not change when the outcome $1$ is not deterministic, as in a CSN, but occurs with certain probability. In this sense, such ``randomized'' SNs do not perform better than CSNs.

The interesting point is that, since there are graphs for which $\omega(G) < \alpha^* (\bar{G})$, then there should exist SNs in which $\mathcal{T}$ goes beyond the maximum value for CSNs.


\section{Quantum social networks}


We shall introduce now a natural SN for which $\mathcal{T}$ may be larger than the maximum for any CSN represented by the same graph. A quantum SN (QSN) is defined as a SN in which each actor $i$ is associated with a quantum state $|\psi_i\rangle$. The states are chosen to reflect the graph of the network in the following sense. Non-adjacent (adjacent) vertices in the graph correspond to orthogonal (non-orthogonal) states. It is always possible to associate quantum states to the actors of any network fulfilling the orthogonality relationships imposed by its graph \cite{LSS89}. Reciprocally, any set of quantum states defines a QSN.

\begin{figure}[t]
 \centerline{\includegraphics[scale=0.35]{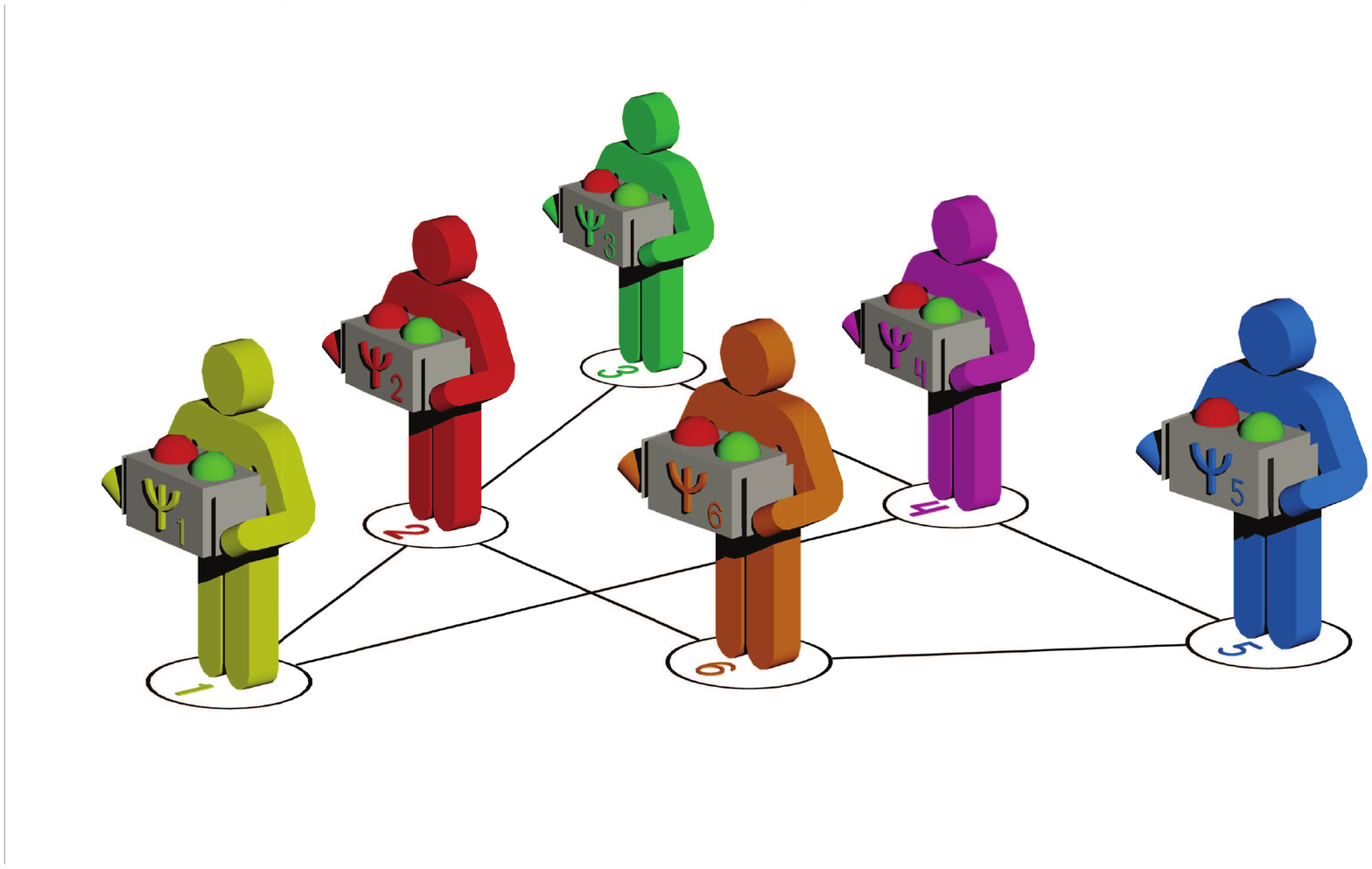}}
 \vspace{-15mm}
 \caption{A QSN can be visualized as a CSN in which each actor $i$ has a device to measure a quantum state $|\psi_i\rangle$.}
 \label{Fig2}
\end{figure}

A QSN can be constructed from a CSN by assigning to actor $i$ a device to test the quantum state $|\psi_i\rangle$, as illustrated in Fig.~\ref{Fig2}. The characterization of the QSN in terms of yes-no tests $T_i$ satisfying rules (I)--(III) is as follows. Each device receives a system $S$ in a quantum state $\rho$ as input, and gives as output either the state $|\psi_i\rangle$ and the outcome 1, or a state orthogonal to $|\psi_i\rangle$ and the outcome 0. These tests are measurements represented in quantum mechanics by rank-1 projectors. Note that projective measurements are the simplest repeatable measurements in quantum mechanics [in agreement with rule (II)], whereas general measurements represented by POVMs are not repeatable.


In a QSN, $\mathcal{T}$ may be larger than the maximum for any CSN represented by the same graph. If each and every actor is provided with the same input state $|\Psi\rangle$, according to quantum mechanics the probability of getting the outcome 1 when performing $T_i$ for a randomly chosen $i$ is now $\mathcal{T}=\frac{1}{n} \sum_{i=1}^{n} |\langle\Psi|\psi_{i}\rangle|^{2}$. Given $G$, the quantity $\frac{1}{n} \max \sum_{i=1}^{n} |\langle\Psi|\psi_{i}\rangle|^{2}$, where the maximum is taken over all quantum vectors $|\Psi\rangle$ and $|\psi_{i}\rangle$ and all dimensions, gives the maximum value of $\mathcal{T}$ for any QSN. This number is equal to $\frac{\vartheta (\bar{G})}{n}$, where $\vartheta(\bar{G})$ is the Lov\'asz number \cite{Lovasz79} of $\bar{G}$, which can be computed to arbitrary precision by semi-definite programming in polynomial time (see the Appendix).

The Lov\'asz number was introduced as an upper bound of the Shannon capacity of a graph \cite{Shannon56}, and it is sandwiched between the clique number $\omega(G)$ and the chromatic number $\chi(G)$ of a graph: $\omega(G) \leq \vartheta(\bar{G}) \leq \chi(G)$ \cite{Knuth94}. The interesting point is that, for those graphs such that $\vartheta(\bar{G}) > \omega(G)$, QSNs outperform CSNs.

On the other hand, $\vartheta(\bar{G})$ is upper bounded by the fractional packing number $\alpha^* (\bar{G})$, as was shown by Lov\'asz in \cite{Lovasz79}. In a nutshell, $\mathcal{T}$ and its three upper bounds fulfil $\mathcal{T} \stackrel{\mbox{\tiny{ CSN}}}{\leq} \frac{\omega(G)}{n} \stackrel{\mbox{\tiny{QSN}}}{\leq} \frac{\vartheta(\bar{G})}{n} \stackrel{\mbox{\tiny{GSN}}}{\leq} \frac{\alpha^* (\bar{G})}{n}$. For example, for the SNs in Figs. \ref{Fig1} and \ref{Fig2}, one has $\mathcal{T} \stackrel{\mbox{\tiny{ CSN}}}{\leq} \frac{1}{3} \stackrel{\mbox{\tiny{QSN}}}{\leq} \frac{\sqrt{5}}{6} \stackrel{\mbox{\tiny{GSN}}}{\leq} \frac{5}{12}$.

These numbers, $\omega(G)$ [which is equal to the independence number $\alpha(\bar{G})$ of the complement graph], $\vartheta(\bar{G})$ and $\alpha^* (\bar{G})$ have previously appeared in quantum information, in the discussion of the quantum channel version of Shannon's zero-error capacity problem \cite{DSW10,CLMW10}, and in foundations of quantum mechanics, in the discussion of non-contextuality inequalities \cite{CSW10}.


\begin{figure}[t]
\begin{center}
\centerline{\hspace{25mm}\includegraphics[scale=0.54]{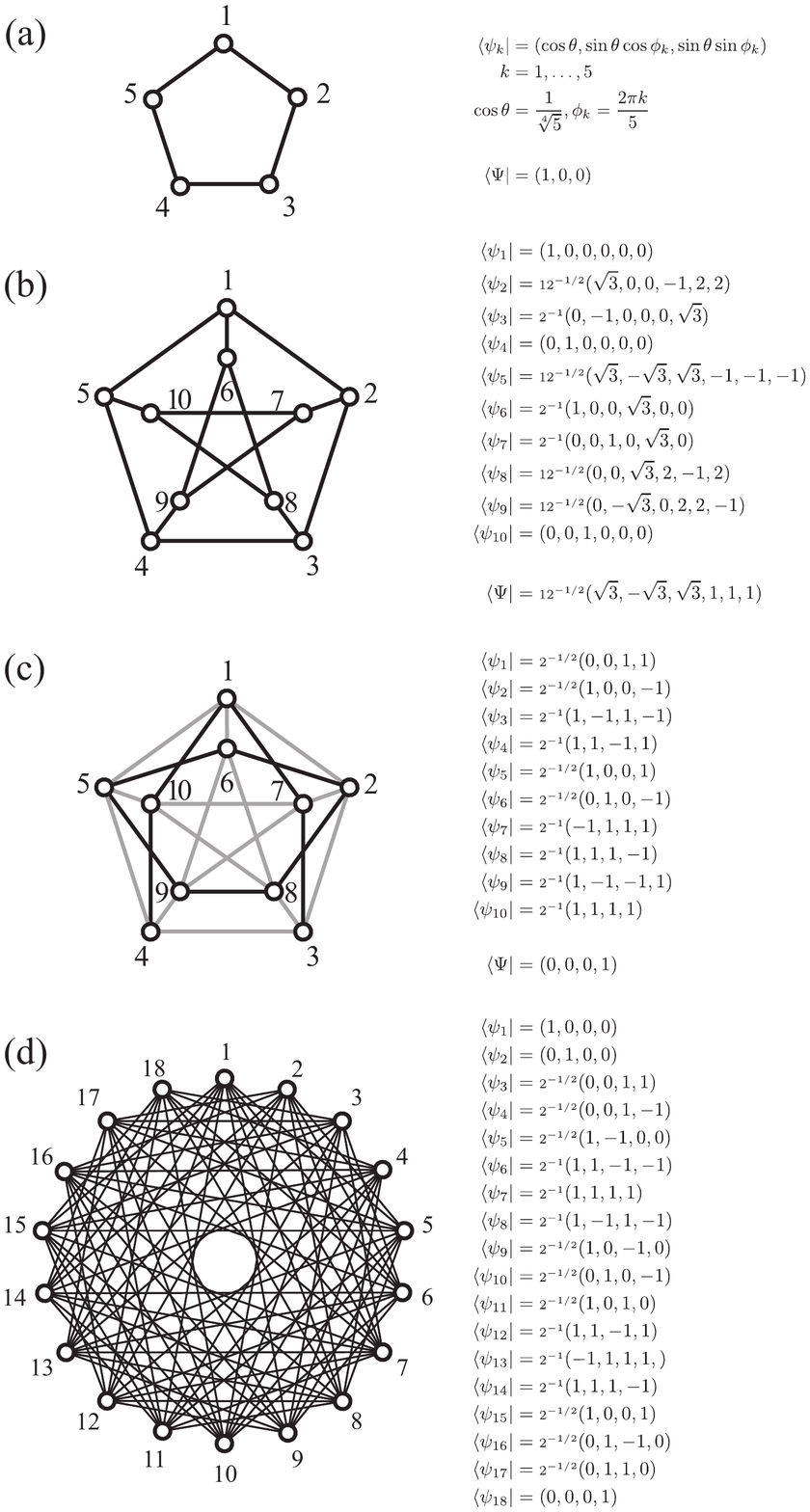}}
\caption{Graphs for which QSNs can outperform CSNs. The right side displays the quantum states $|\psi_i\rangle$ and $|\Psi\rangle$ needed for the maximum quantum advantage using a quantum system of the smallest dimension. (a) is the simplest graph for which QSNs can outperform CSNs. (b) and (c) are the simplest graphs for which QSNs cannot be improved. Graph (c) has the same edges as graph (b) plus extra ones. (d) is the simplest known graph in which the quantum advantage is independent of $|\Psi\rangle$.}
\label{Fig3}
\end{center}
\end{figure}


We generated all non-isomorphic connected graphs with less than 11 vertices (more than $11 \times 10^6$ graphs) and singled out those for which $\vartheta(\bar{G}) > \omega(G)$ (as explained in the Appendix). The graph with less number of vertices such that $\vartheta(\bar{G}) > \omega(G)$ is the pentagon, for which $\omega(G)=2$ and $\vartheta(\bar{G})=\sqrt{5}$, which can be attained using a quantum system of dimension $d=3$ [see Fig.~\ref{Fig3} (a)]. The second simplest graph for which $\vartheta(\bar{G}) > \omega(G)$ is the one in Figs. 1 and 2. For a given number of vertices, the number of graphs such that the maximum of $\mathcal{T}$ for QSNs is larger than for CSNs rapidly increases. The complete list of these graphs with less than 11 vertices is provided in the Appendix.

Interestingly, the probability that the graph of an arbitrary network with a large number of actors contains induced graphs in which a QSN outperforms a CSN is almost identity. This follows from a result in graph theory according to which an arbitrarily large graph contains with almost certainty an induced copy of every graph \cite{Diestel10}. A graph $H$ is said to contain an induced copy of $G$ when $G$ is a subgraph of $H$ obtained by removing some of the vertices and all the edges incident to these vertices.

Moreover, a stronger result can be proven. The probability that QSNs outperform CSNs for an arbitrarily large graph is almost identity. This follows from the observation \cite{FK03} that, while for an $n$-vertex random graph with edges generated with probability $1/2$ the value of $\omega(G)$ is almost surely \cite{AS00} roughly $2 \log_2 n$, the value of $\vartheta(G)$ is almost surely \cite{Juhasz82} $o(\sqrt{n})$.

Once one has identified a graph for which $\vartheta(\bar{G}) > \omega(G)$, one can compute the quantum states $|\Psi\rangle$ and $|\psi_i\rangle$ of minimum dimensionality $\xi(G)$ providing the optimal quantum solution, the one that maximizes $\mathcal{T}$. For the simplest graph with quantum advantage these states are in Fig.~\ref{Fig3} (a). The state $|\Psi\rangle$ is the initial state of $S$ needed to obtain the maximum quantum advantage.


\section{Social networks with no-better-than-quantum advantage}


Remarkably, there are graphs for which QSNs outperform CSNs but no GSN outperforms the best QSN: those satisfying $\omega(G)<\vartheta(\bar{G})=\alpha^* (\bar{G})$. To single out such graphs is particularly interesting because they would allow us to construct the best GSN in a simple way.

We identified all the graphs with less than 11 vertices with $\omega(G)<\vartheta(\bar{G})=\alpha^* (\bar{G})$. There are only four of them. The simplest one is in Fig.~\ref{Fig3} (b). Its quantum realization requires a $d=6$ quantum system (e.g., a qubit-qutrit system) and $\omega(G)=2$ and $\vartheta(\bar{G})=\frac{5}{2}$. The second simplest graph contains the first one, and it is shown in Fig.~\ref{Fig3} (c). It only requires a $d=4$ quantum system; for this graph $\omega(G)=3$ and $\vartheta(\bar{G})=\frac{7}{2}$. The other two graphs are the one in Fig.~\ref{Fig3} (c) with one or two extra edges, as shown in the Appendix.

In all the graphs we have explored so far, the quantum advantage requires the preparation of $S$ in a specific quantum state $|\Psi\rangle$. However, as the complexity of the network increases this requirement becomes unnecessary. This is due to the fact that there are graphs for which the quantum advantage is independent of $|\Psi\rangle$; thus any quantum state (pure or mixed, including maximally mixed) can be used as initial state for the tests $T_i$.

As proven in the Appendix, any set of quantum states $|\psi_{i}\rangle$ belonging to the class of the so-called Kochen-Specker sets \cite{KS67,Peres95} defines a QSN in which the quantum advantage is independent of the state $|\Psi\rangle$. The graph corresponding to the SN associated to the simplest Kochen-Specker set \cite{CEG96} is illustrated in Fig.~\ref{Fig3} (d). It has $\omega(G)=4$ and $\vartheta(\bar{G})=\frac{9}{2}$, requires a $d=4$ quantum system and no GSN can outperform it. Methods to generate Kochen-Specker sets \cite{ZP93,CG96,CEG05,PMMM05} can be used to obtain QSNs with all these features.


\section{Final remarks}


Any actual SN through the internet, like Facebook or Twitter, is complex enough to potentially benefit from assigning quantum tests to the actors. An example is the following: suppose that a company wants to sell a product to as many Facebook users as possible. Under the (correct) assumption that Facebook is a CSN, the optimal strategy would be to identify the biggest subgroup of mutually linked actors, single out their common interest, and then design a commercial targeting this common interest. However, if Facebook were a QSN with exactly the same links as the actual Facebook, then the company would have a larger positive feedback by linking its commercial to the results of the quantum tests.

As in a CSN, the vertices of a QSN can be organized in communities or clusters, with many edges joining vertices of the same cluster and comparatively few edges joining vertices of different clusters. Given a graph $G$, community detection might be simpler if the graph represents a QSN rather than a CSN. The reason is that a QSN with a given $G$ requires a (quantum) physical system of dimension (i.e., number of perfectly distinguishable states) $d_Q=\xi(\bar{G})$, with $\xi(\bar{G})$ the orthogonal rank of the complement of $G$, defined as the minimum $d$ such that there exists an orthogonal representation of $\bar{G}$ in $d$ dimensions (i.e. a function mapping non-adjacent vertices in $G$ to orthogonal vectors in $\mathcal{C}^d$). However, building a CSN requires a physical system of dimension $d_C=i(G)$ (e.g., for the $G$ in Fig. 1, there are $i(G)=7$ distinguishable states $\rho$: jazz,\ldots, running). $d_Q \le d_C$ and, in most cases, $d_Q < d_C$. As an example, while for the graphs in Fig. 3 (a)--(d), $d_C$ is $5,15,10$ and $18$; $d_Q$ is $3,6,4$ and $4$, respectively. Once a community is detected, the study of its induced subgraph will tell us whether or not it has a quantum advantage. Note that QSNs with no global quantum advantage can contain induced subgraphs (e.g., representing communities) with quantum advantage.

On the experimental side, constructing a simple QSN with advantage over its classical counterpart is within actual experimental capabilities. The simplest example is a pentagon in which each actor has a device for testing the appropriate quantum state.


\begin{acknowledgments}
The authors thank I. Herbauts, S. Severini and A. Winter for valuable discussions and C. Santana for graphical support. This work was supported by the Projects No.\ FIS2008-05596, No.\ MTM2008-05866, No.\ FIS2011-29400 and No.\ P06-FQM-01649, the Research Council of Norway and the Wenner-Gren Foundation.\\
\end{acknowledgments}


\section{Appendix}


We explain how we obtained all graphs $G$ with less than 11 vertices for which $\vartheta(\bar{G}) > \omega(G)$.
We also prove that a set of quantum states belonging to the class of Kochen-Specker sets defines a quantum social network in which the quantum advantage is independent of the state.


\subsection{Finding graphs in which QSNs can outperform CSNs}


To obtain all SNs with less than 11 actors in which the assignment of quantum states can outperform the corresponding CSNs, we generated all non-isomorphic connected graphs using {\tt nauty} \cite{McKay90}, and then we calculated $\omega(G)$ (using {\tt Mathematica} \cite{Mathematica}), $\vartheta(\bar{G})$ (using {\tt SeDuMi} \cite{SeDuMi} and also {\tt DSDP} \cite{Benson,BYZ00}) and $\alpha^*(\bar{G})$ (using {\tt Mathematica} from the clique-vertex incidence matrix of $\bar{G}$, obtained from the adjacency matrix of $\bar{G}$ calculated using {\tt MACE} \cite{MACE,MU04} for enumerating all maximal cliques). In addition, we obtained the minimum dimensionality $\xi(G)$ of the quantum system in which the maximum quantum versus classical advantage occurs, or a lower bound of $\xi(G)$, by identifying subgraphs in $\bar{G}$ which are geometrically impossible in a space of lower dimensionality. For example, the simplest impossible graph in dimension $d=1$ consists of two non-linked (non-orthogonal) vertices in $\bar{G}$; in $d=2$, three vertices, one of them linked to the other two. From these two impossible graphs, one can recursively construct impossible graphs in any dimension~$d$ by adding two vertices linked to all vertices of an impossible graph in~$d-2$. For example, if $\bar{G}$ contains a square, then $\xi(G) > 3$. Finally, we have calculated the minimum dimensionality $i(G)$ needed for a CSN by using a program based on {\tt nauty}, {\tt very-nauty} \cite{Briggs} and \cite{KSW78}.


Table \ref{TableI} contains the number of non-isomorphic graphs with a given number of vertices, up to 10 vertices; the number of them in which QSNs outperform CSNs, and for the latter, the number of those for which no GSN outperforms the best QSN. All non-isomorphic graphs with less than 11 vertices (around $10^6$) in which QSNs outperform CSNs are presented in \cite{Web}.


\begin{table}[h]
\begin{center}
\caption{Number of non-isomorphic graphs with vertices ranging from 5 to 10 corresponding to SNs in which the assignment of quantum states to the actors provides advantage, and number of them in which the advantage cannot be improved by GSNs.} \label{TableI}
\vspace{5mm}
\begin{tabular}{c c c c}
Vertices\;\; & Graphs\;\; & \parbox{3cm}{With quantum advantage}\;\; & \parbox{3cm}{With no-better-than-quantum advantage}\\
\hline
$5$ & $21$ & $1$ & $0$ \\
$6$ & $112$ & $2$ & $0$ \\
$7$ & $853$ & $28$& $0$ \\
$8$ & $11117$ & $456$ & $0$ \\
$9$ & $261080$ & $15951$ & $0$ \\
$10$ & $11716571$ & $957639$ & $4$ \\
\end{tabular}
\end{center}
\end{table}

\subsection{State-independent QSNs}

A Kochen-Specker (KS) \cite{KS67} set in dimension $d$ is a set of rays $S$ in the $d$-dimensional complex space such that there is no function $f:S\rightarrow\{0,1\}$ satisfying that for all orthonormal bases $b \subseteq S$, $\sum_{u \in b} f(u)=1$.

\emph{Proposition: }For any KS set in dimension $d$ represented by a graph $G$, $\omega(G) < \vartheta(\bar{G})$ for any initial state in dimension $d$.

\emph{Proof: }For an $n$-ray KS set in dimension $d$, $\vartheta(\bar{G})=\frac{n}{d}$, since $\vartheta(\bar{G})$ is the same for any initial state, including the maximally mixed state $\rho=\frac{1}{d} \openone$ (where $\openone$ represents the identity matrix). $\omega(G)$ cannot reach this number since, by definition of KS set in dimension $d$, there is no way to assign 0 or 1 to their elements in such a way that, for every clique of size $d$ in the complement graph $\bar{G}$, which corresponds to a basis $b \subseteq S$, $d-1$ elements are 0 and one is 1. This means that the best possible assignment respecting that two non-adjacent vertices in $G$ cannot be both 1 includes at least one clique $C$ in $\bar{G}$ for which 0 is assigned to the $d$ elements. The KS set can be expanded so that every vector belongs to a clique of size $d$ in $\bar{G}$, and the assignments can be replicated an integer number $m$ such that $m \omega(G)$ and $m \vartheta(\bar{G})$ can be expressed as a sum of elements grouped in cliques. The contribution of each clique is either 0 or 1. In $m \vartheta(\bar{G})$ all cliques's contribution is 1, whereas in $m \omega(G)$ the contribution of the clique $C$ in which the assignment fails is 0.\hfill \endproof



\end{document}